\documentclass[nofootinbib,twocolumn,pre,letterpaper,longbibliography,superscriptaddress]{revtex4-2}
\usepackage[english]{babel}
\usepackage{amsthm}
\newtheorem{definition}{Definition}
\usepackage{multirow}
\usepackage{latexsym,amsmath,amssymb,amsfonts,graphicx,color,amsthm}
\newtheorem{theorem}{Theorem}
\newtheorem{corollary}{Corollary}[theorem]
\usepackage{enumerate}
\usepackage{braket}
\usepackage{adjustbox}
\usepackage{footnote}
\usepackage[dvipsnames]{xcolor}
\usepackage{tipa}
\usepackage{textcomp}
\usepackage{fontenc}
\usepackage{mathrsfs}
\usepackage{color}
\usepackage{colortbl}
\usepackage{graphics,graphicx}
\usepackage{txfonts}
\usepackage{caption}
\usepackage{subcaption}

\usepackage{amsmath}
\usepackage{amsfonts,amssymb,amstext,amscd,amsthm,bbold}
\usepackage{comment,float}
\newcommand{\tr}{\textcolor{red}}

\usepackage[colorlinks=true,linkcolor=blue,urlcolor=blue,citecolor=blue]{hyperref}

\definecolor{myurlcolor}{rgb}{0,0,0.7}

\usepackage{float}
\usepackage{hyperref}
\def\tr{\operatorname{tr}}

\renewcommand{\o}{\omega}
\newcommand{\pt}{\mathcal{P}\mathcal{T}}        

\newcommand{\h}{\mathcal{H}}
\renewcommand{\ket}[1]{\vert#1\rangle}

\renewcommand{\bra}[1]{\langle#1\vert}
\newcommand{\miniprod}[2]{\langle#1\vert#2\rangle}

\begin{document}
\title{Decomposition of a system in pseudo-Hermitian quantum mechanics}
\author{Himanshu Badhani}
\email{himanshub@imsc.res.in, himanshubadhani@gmail.com}
\affiliation{Optics \& Quantum Information Group, The Institute of Mathematical Sciences, C. I. T. Campus, Taramani, Chennai - 600113, India}
\affiliation{Homi Bhabha National Institute, Training School Complex, Anushakti Nagar, Mumbai 400094, India}
\author{Sibasish Ghosh}
\email{sibasish@imsc.res.in}
\affiliation{Optics \& Quantum Information Group, The Institute of Mathematical Sciences, C. I. T. Campus, Taramani, Chennai - 600113, India}
\affiliation{Homi Bhabha National Institute, Training School Complex, Anushakti Nagar, Mumbai 400094, India}
\begin{abstract}

This work outlines a consistent method of identifying subsystems in finite-dimensional Hilbert spaces, independent of the underlying inner-product structure. Such Hilbert spaces arise in $\pt$-symmetric quantum mechanics, where a non-Hermitian Hamiltonian is made self-adjoint by changing the inner product using the so-called ``metric operator". This is the framework of pseudo-Hermitian quantum mechanics. For composite quantum systems in this framework, defining subsystems is generally considered feasible only when the metric operator is chosen to have a tensor product form so that a partial trace operation can be well defined. In this work, we use arguments from algebraic quantum mechanics to show that the subsystems can be well-defined in every metric space -- irrespective of whether or not the metric is of tensor product form. This is done by identifying subsystems with a decomposition of the underlying $C^*$-algebra into commuting subalgebras. Although the choice of the metric is known to have no effect on the system's statistics, we show that different choices of the metric can lead to inequivalent subsystem decompositions. Each of the subsystems can be tomographically constructed and these subsystems satisfy the no-signalling principle. With these results, we put all the choices of the metric operator on an equal footing for composite systems.
\end{abstract}

\maketitle
\section{Introduction}
It is an axiomatic assumption in quantum theory that the Hilbert space of a composite system has a tensor product decomposition. That is to say, a system with two degrees of freedom, represented by the set of variables $m_1$ and $m_2$ has the wavefunction $\psi(m_1,m_2)\in \mathcal{H}$ in the form $\psi(m_1,m_2)=\sum_i\psi_i(m_1)\otimes\varphi_i(m_2)\in\mathcal{H}_1\otimes\mathcal{H}_2$. Such a decomposition allows for a description of the subsystem \textit{localized} to a subspace of the total Hilbert space, with operations on this subsystem defined as operations on this subspace. While this axiom has a footing in some fundamental principles of functional analysis and can be argued from deeper laws of nature\, \citep{manik22,manik23}, in certain situations, it does not hold good. The most celebrated demonstration of such a deviation is the system of indistinguishable particles\,\citep{bal13,sibasish} and systems with infinite degrees of freedom described by quantum field theory\,\citep{witten,type3}. Another example, and the central topic of this work, is the system that evolves under $\pt$-symmetric or pseudo-Hermitian Hamiltonians \citep{bender,ali03}. Pseudo-Hermitian operators are the non-Hermitian operators with eigenvalues in complex conjugate pairs or real eigenvalues (quasi-Hermitian). They arise naturally in certain optical settings with equal gain and loss\,\citep{nat1,nat2}. It was realized that such evolutions can be seen as unitary evolutions in Hilbert spaces with modified inner product structures induced by a \textit{metric operator} \,\citep{ali02,brody}. In fact, such a description is necessary to avoid fallacious results like the violation of the no-signalling principle\, \citep{lee}.
\\
\\ 
For composite systems under such evolutions, the partial trace operation is well-defined if the metric operator allows a tensor product decomposition of the Hilbert space. Such choices of the metric operators are often chosen over other metric operators to define the subsystem\, \citep{moise,nori}. However, it can so happen that for a given Hamiltonian, none of the choices of the metric operator allow for a tensor product structure\,\citep{hb23}. In such situations, claiming that subsystems cannot be defined would be incorrect, especially if the experimentalist can probe the different degrees of freedom independently of each other. Therefore, to define subsystems in such Hilbert spaces one must use an operational definition of the subsystems, which is the main goal of this article.
\\\\
In this work, we outline the problems that arise in defining a subsystem in pseudo-Hermitian quantum mechanics and resolve them using tools from the algebraic formulation of quantum mechanics. In the literature, the issue of defining subsystems independently of the Hilbert space structure has been explored by decomposing the underlying $C^*$-algebra\,\citep{paolo01,paolo03}. We use this algebraic basis to define subsystems in non-Euclidean inner product spaces. This allows us to put all the choices of the metric operator on equal footing. The central argument of this work is that different decompositions of the underlying algebra correspond to particular families of inner product spaces used to represent the system. Consequently, while a quasi-Hermitian evolution can be represented as unitary evolution in a family of inner product spaces, different choices of the inner product spaces can result in the probing of different subsystem decompositions. While thus far the literature has relied on a genuine tensor product structure of the inner product spaces, we show that under any choice of the metric operator, the no-signalling principle holds for a well-defined subsystem.
\\\\
The article is structured as follows: in this section, we will review the aspects of pseudo-Hermitian quantum mechanics. In section\ \ref{sec:alg} we will give a short introduction to $C^*$-algebras and their pseudo-Hermitian representations. We will define bi-partitions of a system, in a given inner product space and show the dependence of the partitioning on the choice of the inner product in section\ \ref{sec:partition}. In section \ref{sec:tomo} we will show that local tomography can be performed for systems in any metric space, and the no-signalling principle is satisfied. We conclude with a short discussion in section\ \ref{sec:conclusion}.

\subsection{Underlying vector space and the metric}\label{subsec:vecspace}
\noindent
The ``pure" state of a classical system at a given time is given by a set of real numbers $\{m_1,m_2,...\}$ that correspond to a point in the configuration space. In other words, a measurement of the system can yield these values simultaneously.  Let $m_i\in M_i$ where $M_i$ is a set of possible values the state can take. We call a system bipartite if the underlying configuration space is $M=M_1\times M_2$. This means that the experimentalist has the means to measure the parameters of $M_1$ and $M_2$ at the same time. For example, by placing the spin detectors at different points in space, one can measure the spin and position of a system. A quantum state is a complex-valued continuous function over the configuration space, given by $\psi(m_1,m_2,...)$. If the configuration space is a compact subset of $\mathbb{R}^n$ (or more generally a locally compact Hausdorff space) then the space of smooth functions is a vector space $C^\infty(M)$, where one can also define a  ``norm" for the vectors. This defines a Hilbert space. We are generally interested in a subspace $\mathcal{F}(M)\subset C^\infty(M)$. The elements of this space are the function $f(m_1,m_2)$ such that $f(m_1,\cdot)=f_2(\cdot)\in \mathcal{F}_2(M_2)$ and $f(\cdot,m_2)=f_1(\cdot)\in \mathcal{F}_1(M_1)$, where $\mathcal{F}_1(M_1)$ and $\mathcal{F}_2(M_2)$ are set of continuous functions over $M_1$ and $M_2$. Let the set $\mathcal{F}_1(M_1)\otimes \mathcal{F}_2(M_2)$ denote the closure of the set of linear combinations of functions of the form $\varphi(m_1,m_2)=f_1(m_1)\times f_2(m_2)$. Then, we have the equality $\mathcal{F}_{1,2}(M_1\times M_2)\cong\mathcal{F}_1(M_1)\otimes \mathcal{F}_2(M_2)$ if at least one of the sets $\mathcal{F}_i(M_i)$ is a complete normed vector space \, \citep{cart}.
\\
These results from functional analysis lead us to conclude that for a system whose underlying configuration space has a Cartesian product form, the vector space of pure quantum states should have a tensor product decomposition. In this way, one motivates the \textit{tensor product postulate} for composite quantum systems. This vector space is converted to a Hilbert space by introducing an inner product map. As we discuss below, the tensor product decomposition of the resulting Hilbert space depends on one's choice of the inner product map.
\subsection{Overview of pseudo-Hermitian quantum mechanics}\label{subsec:PHoverview}
\noindent
Given two elements $\psi$ and $\phi$ in a vector space $\mathcal{V}$, we define a Euclidean inner product map $\miniprod{\psi}{\phi}\ = \psi^\dagger \phi$, where $\psi^\dagger$ is the conjugate transpose of the element $\psi$. With respect to this inner product, one can define a class of non-Euclidean inner products facilitated by a Hermitian, positive-definite operator $G$ called a metric operator: $\miniprod{\psi}{\phi}_G\ := \psi^\dagger G\phi= \miniprod{\psi}{G\phi}$. This gives us a class of inner product spaces $\h_G:=(\mathcal{V},G)$, where we will denote $\h_{\mathbb{1}}$\ simply as $\h$. It can be shown that if $\h_\mathbb{1}$ is a Hilbert space, then $\h_G$ is also a Hilbert space, i.e. complete with respect to the norm induced by their respective inner product map. If the underlying vector space is of the form $\mathcal{V}_1\otimes\mathcal{V}_2$, then the Hilbert space $\h_G=(\mathcal{V}_1\otimes\mathcal{V}_2,G)$ has a tensor product decomposition $\h_G=\h_{G_1}\otimes\h_{G_2}$ if and only if the metric operator has he form $G=G_1\otimes G_2$ where $G_1$ is metric operator on $\mathcal{V}_1$ and similarly for $G_2$. Given a density matrix on $\h_G$, its subsystem is usually characterized by a completely positive and trace-preserving (CPTP) map to a smaller Hilbert subspace through the partial trace operation. However, such a map is well defined only if the Hilbert space $\h_G$ has a tensor product decomposition of the form $\h_{G_1}\otimes\h_{G_2}$. For Hilbert spaces with a metric operator not in the tensor product form, the partial trace operation is not well defined. In this work, we will outline a method to construct the subsystem statistics for the systems defined on such Hilbert spaces.
\\
Hilbert spaces with non-Euclidean metric operators arise naturally in $\pt$-symmetric quantum mechanics. For example, the evolution of systems under balanced gain and loss can be seen as generated by a non-Hermitian Hamiltonian. Such Hamiltonians are symmetric under the consecutive action of parity and time reversal operations, and can have real eigenvalues. All such Hamiltonians $H$ satisfy a condition called \textit{$G$-pseudo-Hermiticity} given by $H^\dagger G=GH$ for a Hermitian operator $G$. If, for a given Hamiltonian $H$, there exists a positive-definite operator $G$ satisfying the pseudo-Hermiticity condition, then $H$ is guaranteed to have real eigenvalues []. Such operators are called \textit{quasi-Hermitian} operators. It turns out that quasi-Hermitian operators have real expectation values in Hilbert spaces with metric operator $G$, and therefore can be seen as \textit{observables} in $\h_G$ as we show below:
\begin{equation}
\begin{aligned}
    \miniprod{\psi}{H\psi}_G &=\miniprod{\psi}{GH\psi}\\
    &=\miniprod{\psi}{H^\dagger G\psi}\\
    &=\miniprod{H\psi}{ G\psi}\\
    &=\miniprod{H\psi}{\psi}_G
\end{aligned}
\end{equation}
In pseudo-Hermitian quantum mechanics, the physical Hilbert space is a metric space with a metric operator $G\neq\mathbb{1}$. In such a space a $G$-pseudo-Hermitian operator is also a quasi-Hermitian operator and can be treated as an observable.
\\
An operator can be an observable in two different Hilbert spaces since the metric operator satisfying the pseudo-Hermiticity condition for a given operator is not unique \citep{hb23}. Let $\h_{G'}$ and $\h_G$ be two such Hilbert spaces for an observable $\mathcal{O}$, i.e. $\mathcal{O}
^\dagger G=G\mathcal{O}$ , and $\mathcal{O}^\dagger G'=G'\mathcal{O}$. Then, there exists a linear operator $T$ such that $T^\dagger G T=G'$ and $[T,\mathcal{O}]=0$.  It can be shown that the space of all allowed metric operators compatible with Hamiltonian $H$, can be parametrized by the set of real vectors $\vec{\lambda}$\, \citep{ali10}. Given $\{\psi_i\}$ and $\{\phi_j\}$ the left and right eigenvectors of $H$ respectively, (given it has a non-degenerate spectrum), the most general metric is given by $G_\lambda=\sum_i \lambda_i\ket{\psi_i}\bra{\phi_i}$.
\\
A pseudo-Hermitian operator in $\h_G$ can be mapped to a Hermitian operator in $\h$ via the \textit{Hermitisization} map 
\begin{equation}
    \mathcal{O}\rightarrow\mathcal{O}_\eta=\eta\mathcal{O}\eta^{-1},
\end{equation}
where $\eta$ is the unique positive square root of the metric operator $G$. Note that $\eta:\psi\rightarrow \eta\psi$ is an isometry between the spaces $\h_G$ and $\h$. Furthermore, since $G'$ is another possible metric operator, $\mathcal{O}_{\eta'}=\eta'\mathcal{O}\eta'^{-1}$ is another Hermitisization of the observable $\mathcal{O}$.
\\
For quasi-Hermitian Hamiltonians, Mostafazadeh's pseudo-Hermitian quantum mechanics\ \cite{ali10} seems to be the most natural formalism. However,  general pseudo-Hermitian Hamiltonians with complex  eigenvalues leave room for a wider choice of inner product spaces and have been applied to various physical systems\ \cite{baga15,ohlsson20, bender13, wu19}
\subsection{Open-system perspective on pseudo-Hermitian Hamiltonians}\label{subsec:opensys}
Pseudo-Hermitian evolutions in quantum systems have been experimentally exhibited many times in recent years\, \citep{li,tang,xue}.  This is typically achieved by introducing an interacting potential wherein the amplitude and frequency of the gain and loss of the probability current are carefully controlled to maintain the potential's $\pt$-symmetry\, \citep{asboth}. That is, the interacting Hamiltonian is invariant under the action of the parity and time reversal operations. This is equivalent to maintaining a balanced gain and loss of amplitudes. Within certain parameters (referred to as the ``unbroken" $\pt$-symmetry regime) the Hamiltonian has a quasi-Hermitian form\, \citep{ali02}. Due to the presence of interaction terms in the Hamiltonian, such an evolution is sometimes seen as an open system dynamics, even though the effective Hamiltonian has real eigenvalues. A quintessential feature that separates a quasi-Hermitian evolution from a genuinely non-Hermitian (non-unitary) evolution is the revival of complete information of the system throughout the dynamics, measured as probability values or the information back-flow. It has been argued\, \citep{gunther,kwabata} that such a complete revival is due to the presence of a finite-sized entangled partner in the environment of the system, where the Hermitisization operation $\psi\rightarrow\eta\psi$ is seen as an embedding of the system in a larger Hermitian system. As we will elucidate in the subsequent sections, the quasi-Hermitian Hamiltonian, although simulated through interaction terms, induces a unitary evolution.
\\
A large class of open system dynamics follows a Lindblad-type master equation, which can be recast as an evolution under an effective non-Hermitian Hamiltonian and a ``jump-term". Post-selecting on the trajectories without a jump, the evolution of the system can be cast as an evolution under the non-Hermitian Hamiltonian. If the effective Hamiltonian is quasi-Hermitian, then one can show that the Lindblad equation is simply the von Neumann equation for a redefined quasi-Hermitian state, and the jump term vanishes (see \ref{lindblad} for details), effectively reproducing the unitary dynamics. On the other hand, in the ``broken regime" where the Hamiltonian has complex eigenvalues, interpreting the Hamiltonian as a generator of time translation can be tricky\ \cite{ali18}. The compatible operator $G$ is not positive definite in this regime, and as a result the corresponding inner product space is not a well-defined Hilbert space. One can also induce a time-dependent metric operator e.g. \cite{ju22}, or use the Euclidean inner product  \cite{halder23,baga24,hb21} to study such Hamiltonians. In this work we will restrict ourselves to the formalism of pseudo-Hermitian quantum mechanics, where the norm is induced by a non-Euclidean positive-definite inner product. Therefore, the form of the Hamiltonian or the regimes of broken or unbroken phases is not of particular importance to us. However, our results can help understand the properties of the composite systems evolving under such Hamiltonians if treated under the pseudo-Hermitian formalism.

\section{The C*-algebra}\label{sec:alg}
An algebra is a set that is closed under summation, scalar multiplication, and multiplication operations. A C* algebra has an additional operation defined on its elements: a * operation, which is a generalization of the conjugate-transpose operation defined on the set of matrices. We start with a C* algebra $\mathcal{A}$ of operators and a positive, linear functional called a \textit{state} $\o$ (with norm 1) defined on this algebra. The algebra and the state $(\mathcal{A},\o)$ together represent the statistical data of the given experiment on a quantum system, e.g. $A\in\mathcal{A}$ can represent the measurements that can be performed on a system, and the number $\o(A)$ will then represent the expectation value of such operation. 
\\\\ 
\textit{GNS representation:} A faithful representation of the algebra is an *-isomorphism to the set of bounded operators on a Hilbert space 
\begin{equation}
\begin{aligned}
  \pi:\mathcal{A} \rightarrow\pi(\mathcal{A})&\subset\mathcal{B}(\h)\\
  \text{such that }\hspace{1cm}
    \pi(A^*)&=\pi(A)^*.
\end{aligned}
\end{equation} 
This is possible since the set of bounded operators on a complex Hilbert space is also a C*-algebra. The conventional Hilbert space structure of the state and operators can be constructed by the Gelfand–Naimark–Segal or GNS construction, whose details are given in the appendix \ref{app:gnsconstruction}. Given a system $(\mathcal{A},\o)$, a GNS cyclic representation gives a triplet $(\h_\o, \pi_\o, \Phi_\o)$ that represents the Hilbert space, the representation map,  and a vacuum/cyclic vector on $\h_\o$ respectively, such that $\pi_\o(A)\ket{\Phi_\omega}=\ket{\psi_A}$ is a general element of $\h_\o$ and the value of the state is given by 
\begin{equation}
    \o(A)=\bra{\Phi_\o}\pi_\o(A)\ket{\Phi_\omega}_{\h_\o}
\end{equation}
Therefore, the choice of the representation fixes the choice of the inner product. It is also a well-known fact that such a representation is unique up to a unitary equivalence. In the subsequent part of this article, we will assume that given a $C^*$ algebra, one can always construct a faithful cyclic representation of a system $(\mathcal{A},\o)$. The state $\o$ is associated with an operator $\rho_\o$ such that $\o(A)=\tr(\rho_\o\pi_\omega(A))$. In particular, we will demand that the operator $\rho_\o$ is an element of the set $\pi_\o(\mathcal{A})$, in which case it is a unique operator fixed by the faithful representation map $\pi_\o$.
\subsection{Cyclic representations in different metric spaces}
\begin{figure}
    \centering
    \includegraphics[width=0.9\linewidth]{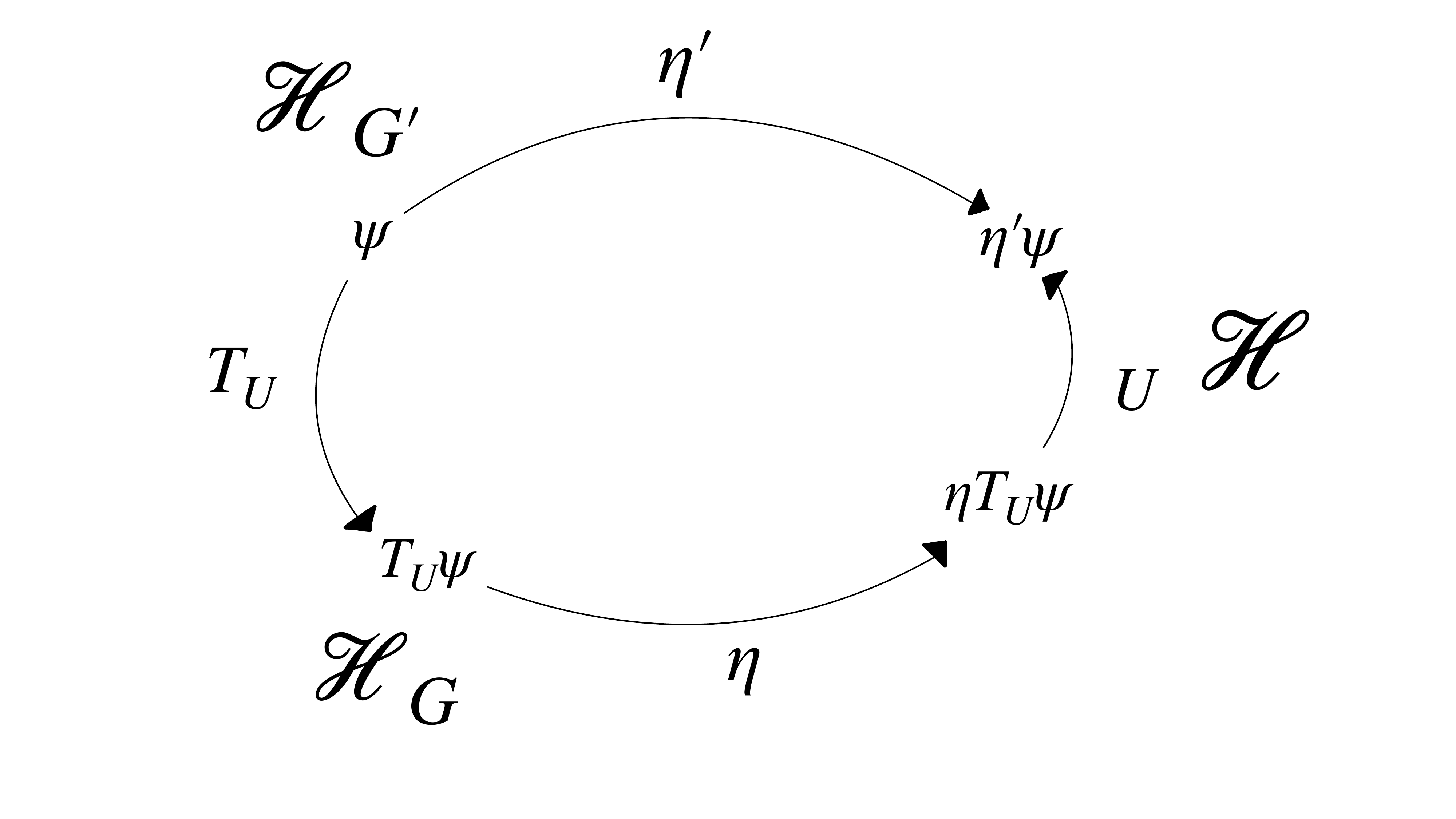}
    \caption{Unitary maps between different metric Hilbert spaces. The state $\psi$ in $\h_{G'}$ can be mapped to a state in $\h$ either through Hermitisization ($\eta'$) or through the Hilbert space $\h_G$ ($\eta T_U$). The resulting states are related by a unitary $U$ in $\h$, which implies $\eta'=U\eta T_U$ }
    \label{fig:unitarymaps}.
\end{figure}

Before exploring the non-Euclidean metric representations of a system, we would like to elucidate the notion of a ``unitary" map between Hilbert spaces with different metric operators. First, we review the notion of isometry, or the norm-preserving map, between metric spaces:
\begin{definition}
\textbf{Isometry:} Given two Hilbert spaces $\mathcal{H}_G$ and $\mathcal{H}_{G'}$, the map $\mathcal{T}:\mathcal{H}_{G'}\rightarrow\mathcal{H}_{G}:\ket{\psi}\rightarrow T\ket{\psi}$ is an isometry if we have $\miniprod{\psi}{\psi}_G=\miniprod{T\psi}{T\psi}_{G'}=\miniprod{\psi}{T^*T\psi}_{G}$. 
\end{definition}
Note that the adjoint of the operator $T$ is defined through the relation $\miniprod{T\psi}{\varphi }_{G}=\miniprod{\psi}{T^*\varphi}_{G'}$. From here one can show that $T^*=G^{-1}T^\dagger G'$ and the map $\mathcal{T}$ is an isometry if and only if  $T^*T=1 \Leftrightarrow T^\dagger G' T=G$.
\begin{equation}
    \begin{aligned}
        \miniprod{T\psi}{\varphi }_{G}&=\miniprod{\psi}{T^*\varphi}_{G'}\\
\Rightarrow \miniprod{T\psi}{G\varphi }&=\miniprod{\psi}{G' T^*\varphi}\\
\Rightarrow   \miniprod{\psi}{T^\dagger G\varphi }&=\miniprod{\psi}{G' T^*\varphi}\\
\Rightarrow T^*&={G'}^{-1}T^\dagger G.
    \end{aligned}
\end{equation}
A unitary map is an isometry which is also invertible. If $\mathcal{T}$ is an invertible map, then $T^{-1}$ exists and $T^*={G'}^{-1}T^\dagger G=T^{-1}$, or equivalently $G'=T^\dagger G T$. Therefore we have the following definition for a unitary between metric spaces:
\begin{definition}
    \textbf{Unitary:} $\mathcal{T}:\mathcal{H}_{G'}\rightarrow\mathcal{H}_{G}:\ket{\psi}\rightarrow T\ket{\psi}$ is a unitary map if and only if  $\mathcal{T}$ is an invertible map and  $G'=T^\dagger GT$.
\end{definition}
The map $\mathcal{O}\to\mathcal{O}'= T^{-1}\mathcal{O}T$ is a unitary map between the set of operators on $\mathcal{B}(\h_G)\to\mathcal{B}\mathcal(\h_{G'})$, and it preserves the expectation value of the observable $\mathcal{O}$ with respect to any state $\psi\in\h_{G'}$ under the unitary $\mathcal{T}$ as is shown below
\begin{equation}
\begin{aligned}
    \miniprod{\psi}{\mathcal{O}'\psi}_{G'}&=\miniprod{\psi}{G'\mathcal{O}'\psi}\\
    &=\miniprod{\psi}{G'T^{-1}\mathcal{O}T\psi}\\
    &=\miniprod{\psi}{T^*G\mathcal{O}T\psi}\\
        &=\miniprod{T\psi}{G\mathcal{O}T\psi}=\miniprod{T\psi}{\mathcal{O}T\psi}_G.
\end{aligned}
\end{equation}
Finally, we note that the Hermitisization operation $\h_G\rightarrow\h:\ket{\psi}\rightarrow \eta\ket{\psi}$ is a special case of such a unitary operation which satisfies $\eta^*=G^{-1}\eta=\eta^{-1}$. Therefore, the Hermitisization map $\mathcal{O}\to\eta\mathcal{O}\eta^{-1}=\mathcal{O}_\eta$ is also a unitary map between the operators on $\h_G$ and $\h$. Note that $\eta T$ and $\eta'$ are two different unitary maps from $\h_{G'}$ to $\h$ (see fig. \ref{fig:unitarymaps}). Therefore, the two states $\eta T\ket{\psi}$ and $\eta'\ket{\psi}$ are related by a unitary on $\h$: $\eta'=U\eta T$. Therefore, the operators $T$ can be parametrized by the unitary $U$ on $\h$, and we denote them by $T_U$. 
\\\\
Armed with this generalized notion of unitary maps, we see that a system represented by $(\mathcal{A},\o)$ can have Euclidean as well as non-Euclidean metric space representation:
\\\\
\textbf{Remark:} The representations $(\mathcal{H},\pi)$, $(\mathcal{H}_G,\pi_G)$ and $(\mathcal{H}_{G'},\pi_{G'})$ are all different cyclic representations of the same system if the representations are related by the following relations for all $A\in\mathcal{A}$
\begin{equation}
    \pi_G(A)=\eta^{-1}\pi(A)\eta=T_U\pi_{G'}(A)T_U^{-1}, \forall\ A\in\mathcal{A}.
\end{equation}
\section{Partitions of a system}\label{sec:partition}
We are now in a position to understand the decomposition of a system into subsystems independent of the chosen representation. The simplest example of a composite quantum system is a pair of distinguishable spin $\dfrac{1}{2}$ particles. In section \ref{sec:example} we will discuss the decomposition of this system in a metric space using the results from this section.
\\\\
Let us have a system $(\mathcal{A},\o)$ with cyclic representation $(H_\o,\pi_\o)$ and two *-subalgebras $\mathcal{A}_1$ and $\mathcal{A}_2$. Let  $\o_1$ and $\o_2$ be the restriction of $\o$ on $ \mathcal{A}_1$ and $\mathcal{A}_2$ respectively. Consider the cyclic representations of the two subsystems $\{\mathcal{A}_1,\o_1\}$ and $\{ \mathcal{A}_2,\o_2\}$ given by $\{\h_{\o_1},\pi^1_{\o_1}\}$ and $\{\h_{\o_2},\pi^2_{\o_2}\}$ respectively. Now, the elements of the form $A_iB_j$, where $A_i\in\mathcal{A}_1$ and $B_j\in\mathcal{A}_2$, can be represented (appendix \ref{app:gnsconstruction} for more details) through the representation $(\h_{\o_1}\otimes\h_{\o_2},\pi^1_{\o_1}\otimes \pi^2_{\o_2})$ where 
\begin{equation}
    \pi^1_{\o_1}\otimes \pi^2_{\o_2}(A_iB_j):=\pi^1_{\o_1}(A_i)\otimes \pi^2_{\o_2}(B_j).
\end{equation}
The set of operators $\{A_iB_j\}$ forms a subset of the algebra $\mathcal{A}_1\vee\mathcal{A}_2$ defined as the minimal algebra containing $\mathcal{A}_1$ and $\mathcal{A}_2$. The two sets coincide if and only if the algebras $\mathcal{A}_1$ and $\mathcal{A}_2$ commute: $[A_i,B_j]=0$ for all $i$ and $j$. In that case $(\h_{\o_1}\otimes\h_{\o_2},\pi^1_{\o_1}\otimes \pi^2_{\o_2})$ is a faithful cyclic representation of the system $\{\mathcal{A}_1\vee\mathcal{A}_2, \o_{12}\}$ , where the state $\o_{12}$ is defined as
\begin{equation}
    \o_{12}(A_iB_j)=\o_1(A_i)\o_2(B_j).
\end{equation}
Finally, if $\mathcal{A}_1\vee\mathcal{A}_2=\mathcal{A}$, then $(\h_{\o_1}\otimes\h_{\o_2},\pi^1_{\o_1}\otimes \pi^2_{\o_2})$ is a faithful representation of $(\mathcal{A},\o)$ where $\h_\o\cong\h_{\o_1}\otimes\h_{\o_2}$ and $\pi_\o(A_iB_j)=\pi^1_{\o_1}(A_i)\otimes\pi^2_{\o_2}(B_j)$. 
\\\\
Therefore, at the level of algebras, the necessary condition for the factorization of the Hilbert space $\h_\o\cong \h_{\o_1}\otimes \h_{\o_2}$ is that there are commuting subalgebras whose sum generates the total algebra. This provides a more operational meaning to the partitioning of a system and is useful in situations where Hilbert space does not factorize for a multipartite system. Motivated by earlier works\, \cite{paolo01}, we provide the following definition for a bi-partition of a quantum system (generalization to a multi-partition is straightforward).
\begin{definition}\label{def1}
\textbf{Bi-partition of a system:}    Given a system $(\mathcal{A},\omega)$, a bi-partition of this system is the pair $\{(\mathcal{A}_1,\o),(\mathcal{A}_2,\o)\}$ if the following conditions are satisfied
\begin{enumerate}
    \item The subalgebras $\mathcal{A}_1$ and $\mathcal{A}_2$ commute: $\forall\, A_i\in \mathcal{A}_1$ and $B_j\in\mathcal{A}_2$, we have $[A_i,B_j]=0$. 
    \item The minimal algebra generated by their union is the algebra $\mathcal{A}$, a relation denoted by the equation $\mathcal{A}_1\vee \mathcal{A}_2=\mathcal{A}$. $\forall\, X\in\mathcal{A}$, $X=\sum_{i,j}A_iB_j$ where $A_i\in \mathcal{A}_1$ and $B_j\in\mathcal{A}_2$. Note that this form of a general element of $\mathcal{A}_1\vee\mathcal{A}_2$ is due to the commutativity of these algebras.
\end{enumerate}
\end{definition}
\noindent
Note that the $C^*$-subalgebras $\mathcal{A}_1$ and $\mathcal{A}_2$ can be replaced by the commuting observable algebras $\mathcal{O}_1$ and $\mathcal{O}_2$. These are the so-called Jordan-Lie-Banach algebras, whose complexification gives us the corresponding $C^*$-algebra\, \cite{landsman}.
\\
\\
Continuing the above discussion, let us suppose that a particular cyclic representation of $(\mathcal{A},\o)$, given by $(\h_\o,\pi_\o)$ allows a \textit{genuine bi-partition} of the system into $(\mathcal{A}_1,\o)$ and $(\mathcal{A}_2,\o)$ and has the metric operator $G$, such that $\h_\o=\h_G=\h_\o^1\otimes\h_\o^2$. One therefore concludes that $G$ has a tensor product structure $G=G_1\otimes G_2$ where $G_1$ and $G_2$ are the metric operators for Hilbert spaces $\h_\o^1$ and $\h_\o^2$, respectively. The corresponding Hermitisization map from such a Hilbert space will maintain the tensor product structure: $\eta=\eta_1\otimes \eta_2:\h_G\rightarrow\h_\mathbb{1}=\h^1\otimes \h^2$. Therefore, \textit{a genuine bi-partition in any metric space is equivalent to a genuine bi-partition in the Euclidean metric Hilbert space.} Hence, we will restrict ourselves to genuine bi-partitions in the Euclidean metric Hilbert space.
\\\\
\textit{Partitioning}: Let us now restrict ourselves to defining subsystems in representations $(\h_G,\pi_G)$, where $G\neq G_1\otimes G_2$, while we do have $\h_\mathbb{1}=\h^1\otimes\h^2$, i.e. the bi-partition in the non-Euclidean metric Hilbert space is not a genuine bi-partition. An observable $\pi_G(A)$ under this representation is also an observable in the Hilbert space $\h_{G'}$ if $\pi_G(A)$ commutes with the unitary map $T_U:\h_{G'}\rightarrow \h_G$. Therefore, given an observable, there can be an infinite number of metric spaces to choose from. However, if we fix a certain set of observables $\{\pi_G(A_i)\}_i$ to probe the system, such that the only operator that commutes with all the elements of the set is a multiple of identity, then the metric space under which this set remains a set of observables is fixed\, \cite{scholtz}. Such a set is called an i\textit{rreducible set of operators} as they do not have any common invariant subspace. This discussion points to the relationship between the choice of the metric operator and the choice of observables. The question then naturally arises, given a partitioning of an algebra, to what extent is the metric operator fixed? The following theorem answers this question.
\begin{theorem}
For a system $\{ \mathcal{A},\o\}$, any two cyclic representations $\pi$ and $\pi'$, with representation spaces $\h_G$ and $\h_{G'}$ respectively, correspond to the same bi-partitioning into system 1 and 2 if and only if they belong to the equivalence class given by:
\begin{equation}\label{repeq}
    [\pi]\equiv\{\pi'(\cdot)=T_U^{-1}\pi(\cdot)T_U; \,  \eta'T_U^{-1}=(U_1\otimes U_2) \eta\}.
    \end{equation} 
    Here $U_1$ and $U_2$ are unitary operators corresponding to the two subsystems. 
\end{theorem}
\begin{proof}
Let
\begin{equation}
    \begin{aligned}
        \mathcal{A}_\pi^1&=\{A\in \mathcal{A}:\eta\pi(A)\eta^{-1}= \mathcal{O}_A\otimes \mathbb{1}\}\\
        \mathcal{A}_{\pi'}^1&=\{A\in \mathcal{A}:\eta'\pi_{G'}(A)\eta'^{-1}= \mathcal{O}_A'\otimes \mathbb{1}\}
    \end{aligned}
\end{equation}
be the two subalgebras of $\mathcal{A}$ defined w.r.t. two different metric operators $G=\eta^2$ and $G'={\eta'}^2$ respectively. One defines the complementary subalgebras $\mathcal{A}_\pi^2=\{A\in \mathcal{A}:\eta\pi_{G}(A)\eta^{-1}= \mathbb{1}\otimes \mathcal{O}_A\}$ and similarly, $\mathcal{A}_{\pi'}^2$. This gives us two bi-partitions of the algebra $\mathcal{A}$ into $\mathcal{A}^1_\pi\vee \mathcal{A}^2_\pi$ and $\mathcal{A}^1_{\pi'}\vee \mathcal{A}^2_{\pi'}$. 
\\\\
Now we will determine the condition for the equality of the two algebras $\mathcal{A}_\pi^1\cong\mathcal{A}_{\pi'}^1$. Let $X\in\mathcal{A}_\pi^1$. We know that the operators under the two Hilbert spaces are related by $\pi'(X)=T_U^{-1}\pi(X)T_U$ such that $\eta'T_U^{-1}=U\eta$. Combining these two relations we have 
\begin{equation}
\eta'\pi_{G'}(X)\eta'^{-1}=U\eta\pi_{G}(X)\eta^{-1}U^\dagger    
\end{equation}
Therefore $X\in \mathcal{A}_1'$ if and only if
\begin{equation}
U \mathcal{O}_X\otimes \mathbb{1}U^\dagger= \mathcal{O}_X'\otimes \mathbb{1}    
\end{equation}
for the given unitary $U$. Similarly we have $\mathcal{A}_\pi^2\cong\mathcal{A}_{\pi'}^2$ if and only if $U (\mathbb{1}\otimes \mathcal{O}_Y) U^\dagger= \mathbb{1}\otimes\mathcal{O}_Y'$ for all $Y\in\mathcal{A}^2_{\pi'}$. Unitaries satisfying these conditions must always be of the form $U=U_1\otimes U_2$ (see e.g. \cite{hulpke06,johnston11}). Therefore, the two partitions induced by the representations $\pi$ and $\pi'$ are equivalent if and only if the intertwining isometry $T_U$ satisfies the relation $\eta'T_U=U_1\otimes U_2\eta$. It is straightforward to see that representations related by such an intertwiner form an equivalence class.
\end{proof}
\noindent
Note that taking $T_U=\eta^{-1}$ gives the Hermitisization of the representation in $\h_G$ and the resulting representation is in the same equivalence class. Therefore, the entanglement structure of the state $\bar{\rho}\in\h_G$ will be the same as that of the state $\rho_\eta=\eta\bar{\rho}\eta^{-1}\in\h$. We will exhibit this fact more clearly in the next section when we discuss the tomographic construction of a composite system in non-Euclidean metric space.
\\\\
In a typical experiment, the pseudo-Hermitian Hamiltonian is simulated through post-selection or equal gain-loss mechanisms. The form of the Hamiltonian is therefore fixed by the experimentalist, and hence one is restricted to a class of metric spaces compatible with this Hamiltonian. Such metric spaces are fixed by the isometry $T_U$, or equivalently, the unitary $U$. As a simple corollary of the above theorem, the partitioning of the system is then fixed by the choice of the Hamiltonian compatible metric operator.
\begin{corollary}\label{cor1}
Given a quasi-Hermitian operator $H$ such that $H^\dagger G=GH$, the choices of the compatible representation spaces $\h_{G'}$ with equivalent partitioning are those belonging to the equivalence class [G] defined by
\begin{equation}
    [G]\equiv\{G'=T_U^\dagger G T_U;\,  \eta'T_{U}^{-1}=(U_1\otimes U_2) \eta\, \text{ and }\ [T_U,H]=0 \}.
    \end{equation}
\end{corollary}
\begin{proof}
Given two cyclic representations $(\h_G,\pi)$ and $(\h_{G'},\pi')$, such that $H=\pi(X)=\pi'(X)$, therefore the representation spaces are $H$-compatible.  There exists an isometry $T_U$ between the Hilbert spaces such that $\pi=T_U\pi'T_U^{-1}$ which implies $[\pi(X),T_U]=0$. From the previous theorem, the two subalgebra decompositions $\mathcal{A}_\pi^1\vee \mathcal{A}_\pi^2$  and $\mathcal{A}_{\pi'}^1\vee \mathcal{A}_{\pi'}^2$ are equivalent if and only if $\eta'T_{U}^{-1}=(U_1\otimes U_2) \eta$.
\end{proof}
As a consequence of the above results, we see that while the choice of the metric does not influence the system's properties, viz. the expectation values of the observables, it can change the properties of the subsystem, e.g. the entanglement between the subsystems. This is not unlike in the case of global unitary rotation on the system, which is simply a choice of a global reference frame, which does not change the expectation values on the system, but does change the subsystems' properties like the entanglement between them. On the other hand, a local unitary transformation keeps the subsystem properties intact. The above theorem explains this observation from an algebraic point of view: performing a local unitary transformation in the Euclidean metric Hilbert space is the same as choosing a metric space from an equivalence class of metric operators, and therefore keeps the underlying algebraic partitioning the same.
\\\\
\textbf{Tensor product structure on $\h_G$: }A tensor product structure (TPS)\, \cite{paolo01} on a Hilbert space $\mathcal{H}_G$ is an isomorphism $\varphi:\mathcal{H}_G\rightarrow\mathcal{H}=\mathcal{H}_1\otimes\mathcal{H}_2$ where $\dim(\mathcal{H}_1)\times\dim(\mathcal{H}_2)=\dim(\mathcal{H}_G)$. We know that this isomorphism given by the map\, $\varphi:\, \ket{\psi}\to \eta\ket{\psi}$. Consider another Hilbert space $\mathcal{H}_{G'}$ with the metric $G'$ and a TPS $\varphi' :\mathcal{H}_{G'}\rightarrow\mathcal{H}=\mathcal{H}_1\otimes\mathcal{H}_2:\psi\rightarrow \eta'\psi$. We have also seen that there is an isometry between the two Hilbert spaces, as defined in previous sections, is $\mathcal{T}:\mathcal{H}_{G'}\rightarrow\mathcal{H}_{G}:\, \psi\rightarrow T_U\ket{\psi}$. Therefore, the map $\varphi'\circ\mathcal{T}^{-1}:\psi\rightarrow \eta'T_U^{-1}\psi$ defines another TPS over the Hilbert space $\mathcal{H}_G$. This gives us a family of TPS on $\h_G$ parametrized by the operator $\eta'$, or equivalently by $U$. If the unitaries parameterizing the TPS are local transformations, i.e. of the form $U=U_1\otimes U_2$, we say they generate equivalent TPS\footnote{Following the terminology of\, \cite{paolo01}}. From corollary \ref{cor1} we see that:
\\[2mm]
\textbf{Remark:} \textit{a family of equivalent TPS on $\h_G$ corresponds to the equivalence class of metric operators $[G]\equiv\{G'=T_U^\dagger G T_U;\,  \eta'T_{U}^{-1}=(U_1\otimes U_2) \eta\},$  and consequently the same partitioning of the total algebra.}
\section{Tomography and the no-signalling principle}\label{sec:tomo}
In this section, we will show that the subsystems can be constructed using usual state tomography on the system. First, we will show that the states can be constructed in $\h_G$ as well as $\h$ using the same tomographic data.

A tomographically complete set of observables, given by $\{A_i\}\subset\mathcal{A}$, satisfies the condition: $\o_1(A_i)=\o_2(A_i)\iff \o_1=\o_2$ for all states $\o_1$ and $\o_2$ on $\mathcal{A}$. Therefore, such a set completely characterizes any given state $\o$. We will show that the corresponding density matrices, $\bar{\rho}\in\mathcal{H}_G$, and $\rho_{\eta}\in\mathcal{H}$ are \textit{tomographically equivalent} i.e., based on the distribution of $\o$ on a tomographically complete set of observables, the experimentalist can construct either of the states. The density matrix for the state, under the representation $(\mathcal{H},\pi)$, is constructed through a sampling operator $W$\, \cite{ohliger} which is given as
\begin{equation}\label{sampling}
W(\rho)=\int d\mu(\pi(A))\miniprod{\pi(A)}{\, \rho}\pi(A).	
\end{equation}
Here $\mu$ is a probability measure on the set of observables $\pi(\mathcal{A})\cong\mathcal{B}(\mathcal{H})$ and $\miniprod{\mathcal{O}_i}{\, \rho}=\tr(O_i^\dagger \rho)=\o(A_i)$ is the Hilbert-Schmidt inner product between the operators $\mathcal{O}_i=\pi(A_i)$ and $\rho$. A faithful reconstruction of the state requires that the operator $W$ be a full-rank operator, which is a constraint on the measure $\mu$. $A$ measure that samples tomographically complete set of observables would produce a full rank sampling operator, in which case the measure is called a tomographically complete measure.
\\\\
Note that given a tomographically complete set of observables $\{\mathcal{O}_i\}$ in the Hilbert space $\mathcal{H}$, the set of observables $\{\bar{\mathcal{O}}_i\}$, where $\bar{\mathcal{O}_i}=\eta^{-1}\mathcal{O}_i\eta=\pi_G(A_i)$, is a tomographically complete set in $\mathcal{H}_G$, due to the fact that $\o(A_i)$ is independent of the cyclic representation. Furthermore, since the norm in the two Hilbert spaces is related by $||\mathcal{O}||=||\bar{\mathcal{O}}||_G$, the measure $\mu_G$ on the space $\mathcal{B}(\mathcal{H}_G)$, defined by $d\mu_G(\bar{\mathcal{O}})=d\mu(\mathcal{O})$, is also a probability measure. Combining these observations we can construct the sampling operator in $\mathcal{B}(\mathcal{H}_G)$:
\begin{equation}\label{samplinginhg}
\begin{aligned}
\overline{W}(\bar{\rho})&=\int d\mu_G(\pi_G(A))\miniprod{\pi_G(A)}{\, \bar{\rho}}_G\pi_G(A)\\
&=\eta^{-1}W(\rho)\eta    
\end{aligned}
\end{equation}
Therefore, if the measure $\mu$ gives a tomographically complete set of observables in $\mathcal{B}(\mathcal{H})$, then $\mu_G$ is also tomographically complete. Furthermore, if $\mu$ is a \textit{tight frame} (i.e. if $W=\mathbb{1}$), then $\mu_G$ is also a tight frame. This means that given the state $\o$ on $\{A_i\}\subset \mathcal{A}$, one can construct the state $\bar{\rho}$ or the state $\rho_\eta=\eta^{-1}\bar{\rho}\eta$  for any invertible, Hermitian operator $\eta$. In this sense, we call $\bar{\rho}$ and $\rho_\eta$ the tomographically equivalent pair of states.
\\\\
In an experiment, the knowledge of a state is given in terms of the statistical data on the observables, represented as the set of possible outcomes $\{m\}$ and the corresponding probability of the outcome when the observable $A_i$ is measured $\equiv\{m,p(m|A_i)\}$. The connection between this data and the equation \eqref{sampling} is established by the Born's rule which tells us that the inner product $\miniprod{\pi(A_i)}{\rho}$ is the expectation value of the measurement results of the observable $A_i$: $\sum_m mp(m|A_i)=\miniprod{\pi(A_i)}{\rho}=\o(A_i)$. From our discussion in the previous sections, we know that this quantity is independent of the chosen cyclic representation of the system $(\mathcal{A},\o)$ i.e. $\miniprod{\pi(A)}{\, \rho}=\miniprod{\pi_G(A)}{\, \bar{\rho}}_G=\o(A)$.
\\\\
\subsection{An example}\label{sec:example}
Take the example of the spin 1/2 system. The set $\{\mathbb{1},\vec{\sigma}\}$ forms a tight frame in the Euclidean metric Hilbert space $\mathcal{H}=\mathbb{C}^2$ since one can write
\begin{equation}
    W(\rho)=\rho=\dfrac{1}{2}\sum_{\mu=0}^1 \tr(\sigma_\mu\rho)\sigma_\mu.
\end{equation}
Since the Pauli matrices form the orthogonal basis for the space of observables in two dimensions, i.e. $\tr(\sigma_i\sigma_j)=2\delta_{ij}$, they correspond to measurements of the spin in three orthogonal directions. A general observable is therefore given by $\vec{\sigma}.\hat{n}$ and is associated with a unit direction $\hat{n}$ in a 3-dimensional real vector space. Let $p(m| \hat{n}_i)$ be the probability of getting the outcome $m$, given the spin was measured in the $\hat{n}_i$ direction, corresponding to the measurement of the $\sigma_i$ observable. By controlling the direction and magnitude of the magnetic field in the Stern-Gerlach setting (see appendix\ \ref{spintomo} for details of spin tomography), one produces the experimental data $\{m,p(m|\hat{n}_i)\}$, and constructs the density matrix: $\rho=\mathbb{1}/2+\sum_{m,i} mp(m|\hat{n}_i)\sigma_i $. 
\\\\
Note that while associating the operator $\sigma_i$ with the direction $\hat{n}_i$, the experimentalist makes \textit{two choices}. The first is the choice of a tight frame from the family of unitarily equivalent tight frames. Since the set of operators $\{\mathbb{1},U\sigma_iU^\dagger\}$ also forms a tight frame in $\h$, the density matrix thus constructed is fixed only up to a unitary equivalence. The second choice is that of the metric space in which the operators $\{\sigma_i\}$ and their unitary equivalents are observables (in this case it is the Euclidean metric space). Equivalently, one can choose the set of operators $\{\bar{\sigma}_i=\eta^{-1}\sigma_i\eta\}$ which are the observables in the metric space $\h_G$ with metric $G=\eta^2$. The set $\{\bar{\sigma}_i\}$ also forms a tight frame and their measurements can be associated with measurements in orthogonal directions since these operators are orthogonal with respect to the Hilbert-Schmidt inner product on $\h_G$: $\tr(\bar{\sigma}^*_i\bar{\sigma}_j)=2\delta_{ij}$. With this choice of the metric space, the resultant operator $\bar{\rho}=\mathbb{1}/2+\sum_{m,i} mp(m|n_i)\bar{\sigma}_i $ is a density matrix in $\h_G$. 
\\\\
Now, consider a composite system of two distinguishable spin-1/2 particles. In the Euclidean metric space $\h=\h_1\otimes\h_2$ the set of operators $\mathcal{A}_{12}:=\{\mathbb{1}_{12},\sigma_i\otimes \sigma_j,\mathbb{1}_1\otimes\sigma_i,\sigma_i\otimes\mathbb{1}_2\}$ forms the tight frame. The experimental setup consists of two independent Stern-Gerlach apparatus mentioned above. We then collect the data to construct the probabilities $p(m_1,m_2|\hat{n}^1_i,\hat{n}^2_j)$ of obtaining the result $m_1$ as a result of measurement in $\hat{n}^1_i$ direction on particle 1 and $m_2$ as a result of measurement in $\hat{n}^2_j$ direction on particle 2. If the two particles are accessible to two distinct parties, namely Alice and Bob (whose system indices will be labeled as $1$ and $2$ respectively), then Alice only has the data $\{m_1,p(m_1|\hat{n}^1_i)\}$ which is a subset of the complete set of data of the system. Can we say that this subset of the complete data describes a subsystem? Alice can construct a state in $\h_1$ namely $\rho_1$ using the Born rule $\sum_{m_1} m_1p(m_1|\hat{n}^1_i)=\tr(\rho_1\sigma_i)$. From the global point of view, this expectation value is represented as $\sum_{m_1} m_1p(m_1|\hat{n}^1_i)=\tr(\rho_{12}\sigma_i\otimes \mathbb{1})$. Therefore, Alice's data corresponds to the set of observables $\{\sigma_i\otimes\mathbb{1}\}$, which forms a subalgebra of the tight frame $\mathcal{A}_{12}$. With the set $\{\mathbb{1}\otimes\sigma_i\}$, corresponding to the data $\{m_2,p(m_2|\hat{n}_j^2)\}$ belonging to Bob, forming the complementary subalgebra of $\mathcal{A}_{12}$, one can say that this data completely characterizes Alice's subsystem, in this case, represented by the density matrix $\rho_1$.
\\\\
Consider the non-Euclidean metric representation of the system of two particles in $\h_G\neq\h_{G_1}\otimes\h_{G_2}$, as the state $\bar{\rho}_{12}$ related to the density matrix defined above through $\bar{\rho}_{12}=\eta^{-1}\rho_{12}\eta$. Alice, still in possession of the data $\{m,p(m,\hat{n}^1_i)\}$, might be aware of the global metric structure $G$, but due to the structure of the metric, she does not have a notion of local observables. Therefore, she cannot construct a density matrix for particle 1 alone which can be seen as a reduced state of $\bar{\rho}_{12}$. Nevertheless, from the global point of view, her data can be given as 
\begin{equation}
\sum_{m_1} m_1p(m_1|\hat{n}^1_i)=\dfrac{1}{2}\tr(\bar{\rho}_{12}\eta^{-1}(\sigma_i\otimes \mathbb{1})\eta).    
\end{equation}
 Therefore, Alice's data also corresponds to the set of observables $\{\eta^{-1}(\sigma_i\otimes\mathbb{1})\eta\}$, which forms a subalgebra of the total algebra $\eta^{-1}\mathcal{A}\eta$. With $\{\eta^{-1}(\mathbb{1}\otimes \sigma_i)\eta\}$ forming the complementary subalgebra together generating the total $C^*$ algebra $\mathcal{B}(\mathcal{H}_G)$, Alice's data corresponds to a well-defined subsystem of $\h_G$. But as we have established, the data $\{m,p(m|\hat{n}_i^1)\}$ is faithfully represented by $\rho_1=\tr_2(\rho_{12})\in\mathcal{H}_1$. These arguments can be generalized for an arbitrary finite-dimensional system, and therefore, we have the following theorem:
\begin{theorem}
Given a density matrix $\bar{\rho}_{12}$ in $\h_G$, the density matrix $\rho_1=\tr_2(\eta\bar{\rho}_{12}\eta^{-1}$) represents a well-defined subsystem of the system represented by $\bar{\rho}_{12}$ if $\eta\bar{\rho}\eta^{-1}\in\h_1\otimes\h_2$.
\end{theorem}
\noindent
As we discussed, the Hermitisization of the state to a Hermitian density matrix corresponds to a unitarily equivalent representation of the state. Therefore, the bipartite states $\rho_{12}$ and $\bar{\rho}_{12}$ have the same entanglement structure. For instance, if the state $\bar{\rho}_{12}$ (or equivalently the state $\rho_{12}$)  is a pure state, the entropy of the reduced state $\rho_1$ is a measure of entanglement between the subsystems 1 and 2 in both representations.
\\\\
\subsection{The no-signalling principle}\label{sec:nosignalling}  
Suppose there is a constraint on the composite system shared between Alice and Bob, that Alice cannot signal to Bob. This can happen due to the constraints of the experimental setup, e.g. if Alice and Bob's measurement events are space-like separated, or if Alice is in the causal future of Bob. In such a scenario, the no-signalling principle states that the statistics of the subsystem measured by Alice, denoted by $\{m_1,p(m_1|\hat{n}^1_i)\}$, should not change by a measurement on another subsystem belonging to Bob. Taking the example of the composite spin-1/2 system that we just discussed, no-signalling from Alice to Bob means that the conditional probabilities $p(m_1|\hat{n}^1_i)$ must be independent of Bob's measurement directions $\hat{n}_j^2$: 
\begin{equation}
    \sum_{m_2} p(m_1,m_2|\hat{n}^1_i,\hat{n}^2_j)=p(m_1|\hat{n}^1_i)
\end{equation}
Suppose Alice's and Bob's subsystems correspond to the subalgebra $\mathcal{A}_1$ and $\mathcal{A}_2$ respectively. The POVM elements $\{M_j^m\}$ corresponding to the measurements by Bob in the $\hat{n}_j$ direction yielding the result $m$, belong to the subalgebra $\mathcal{A}_2$. Given a representation of the total system in $(\h_G,\pi_G)$, the POVM elements of Bob given by $\pi_G(M_j^m)$ satisfy the relation  $\eta\pi_G(M_j^m)\eta^{-1}=\mathbb{1}_1\otimes\Pi_j^m$,  where $\Pi_j^m$ are the POVMs elements of Bob as represented as operators on $\h_2$, such that $\sum_m\Pi_j^m=\mathbb{1}_2$.  Therefore, for their shared system $\bar{\rho}_{12}$ in the metric space $(\h_G,\pi_G)$,  we have the following relation:
\begin{equation}\label{nosignal}
\begin{aligned}
    \rho_1=\tr_2(\eta\bar{\rho}_{12}\eta^{-1})&=\sum_m \tr_2(\eta \pi_G(M_j^m)\bar{\rho}_{12}\eta^{-1})
\end{aligned}
\end{equation}
The second equality comes from the no-signalling principle in Euclidean metric space stated as the following :
\begin{equation}
\sum_m \tr_2(\eta\pi_G(M_i^m)\bar{\rho}_{12}\eta^{-1})=\sum_m \tr_2(\mathbb{1}\otimes \Pi_i^m\rho_{12})=\tr_2(\rho_{12})    
\end{equation}
Since we have already established that the state $\rho_1$ faithfully represents Alice's subsystem, we can conclude that the no-signalling principle for a bi-partition, as given in the definition \ref{def1}, is obeyed in every metric space. In the literature, one typically shows that the no-signalling principle is satisfied when the metric is separable $G=G_1\otimes G_2$, e.g.\,\cite{nori}, which is only a specific case of the more general principle that we have exhibited here through equation\,\eqref{nosignal}.
\section{Conclusion}\label{sec:conclusion}
 The axiom of the tensor product decomposition of the Hilbert space of composite quantum systems restricts how we understand a subsystem. In a vast family of quantum systems, such a decomposition is not possible, and consequently, a partial trace operation is not well defined. This issue arises in the non-trivial metric space representation of composite systems, which forms a natural description of the systems undergoing balanced gain and loss under the pseudo-Hermitian Hamiltonian. In this work, we have used algebraic tools to provide an operational definition of subsystems in such Hilbert spaces with non-trivial metric structures. We show that given an algebra and a state over it, one can equivalently construct a Hermitian or a pseudo-Hermitian cyclic representation, which are unitarily equivalent if we use the term \textit{unitary} in its most general sense. While, given a pseudo-Hermitian Hamiltonian, one can choose from a family of Hamiltonian compatible metric spaces, we show that different choices of the metric operator can correspond to choosing different system decompositions. Since choosing a metric is equivalent to choosing a reference frame for the system, it makes sense that choosing two different metric representations changes the subsystem properties. Furthermore, the choice of subsystem decomposition is equivalent to a choice of an equivalence class of metric spaces. Finally, we show that from an experimentalist's point of view, local tomography on a subsystem undergoing pseudo-Hermitian evolution creates no paradoxes, and allows one to construct a local description of the subsystem in trivial metric space. The no-signalling principle, which so far has been restricted to tensor product decomposable Hilbert spaces, can be shown to hold good for an algebraically defined subsystem, even when the Hilbert spaces do not have a genuine tensor product structure.
\\
After the mathematical description of pseudo-Hermitian quantum mechanics as a trivial extension of normal quantum mechanics was established, most studies dealing with composite systems in this field have been wary of using the non-separable metric structure due to the ill-defined partial trace operation in such scenarios. From this work, it is clear that such metric operators are as physically meaningful as the separable ones, and there is no clear physical reasoning for choosing a separable metric operator over the non-separable operators to define the subsystems. We hope that this work will be helpful towards a better understanding of composite quantum systems evolving under the $\pt$-symmetric non-Hermitian Hamiltonians.
\\[2mm]
\textbf{Acknowledgments:} H.B. thanks A P Balachandran for useful discussions and Namit Anand for his valuable comments on the draft.

\appendix
\section{Lindbladian dynamics in $\h_G$}\label{lindblad}
Given the Lindblad dynamics of the state $\rho$
\begin{equation}
    \dot{\rho}(t)=-i[H,\rho(t)]+\sum_j-\dfrac{1}{2}\{L_j^\dagger L_j,\rho(t)\}+L_j\rho L_j^\dagger,
\end{equation} 
define $H_e=H-\dfrac{i}{2}\sum_i L_j^\dagger L_j$ (effective non-Hermitian Hamiltonian) so that the above equation can be recast in the following form 
\begin{equation}
    \dot{\rho}(t)=-i(H_e\rho(t)-\rho(t) H_e^\dagger)+\sum_j L_j\rho(t) L_j^\dagger.
\end{equation} 
This shows that the Lindblad equation can be written as an evolution under a non-Hermitian Hamiltonian and an extra term called a ``jump term". Define $M_j=L_j\sqrt{dt}$ and $M_0=\mathbb{1}-iH_e dt$, where $dt$ is an infinitesimal amount of time. Then the following infinitesimal CP map corresponds to the above Lindblad equation:
\begin{equation}
    \Lambda_{t,t+dt}(\rho(t))=\rho(t+dt)=M_0\rho(t) M_0^\dagger+\sum_j M_j\rho(t) M_j^\dagger.
\end{equation}
The probability of the jump in this infinitesimal time interval $t$ to $t+dt$ is given by $dp=\sum_j \tr(L_j^\dagger L_j\rho(t))dt$ and the probability of no jump in this interval is $p_0=1-dp=\tr(M_0^\dagger M_0\rho(t))$. \\
If the effective Hamiltonian is quasi-Hermitian ($H_e^\dagger\eta^2=\eta^2 H_e $), then, the transformation $\rho\rightarrow \rho\eta^2=\bar{\rho}$ rewrites the Lindblad equation as
\begin{equation}
    \dot{\bar{\rho}}(t)=-i[H_e,\bar{\rho}(t)]+\sum_j L_j\bar{\rho}(t) L_j^*
\end{equation}
Here $A^*:=\eta^{-2}A^\dagger \eta^2$. This corresponds to the following stochastic evolution in $\h_G$
\begin{equation}
    \bar{\rho}(t+dt)=\bar{\Lambda}_{t,t+dt}(\bar{\rho}(t))=M_0\bar{\rho}(t) M_0^*+\sum_j M_j\bar{\rho}(t) M_j^*.
\end{equation}
Now the jump probability within the time interval $t$ to $t+dt$ is $dp=\sum_j \tr(M_j^*M_j\bar{\rho})=\sum_j \tr(L_j^\dagger \eta^2 L_j \rho)dt$ and the probability of no jump is $p_0=\tr(M_0^* M_0\bar{\rho})$. However, one can check that $M_0^*M_0=\mathbb{1}$, therefore, $p_0=\tr(\bar{\rho})$. Since one can always choose $\eta^2$ s.t. $\tr(\bar{\rho})=1$, the quasi-Hermiticity condition tells us $\sum_j \tr(L_j^\dagger L_j\eta^2)=0$. From here we can conclude $\sum_j \tr(L_j^\dagger \eta^2 L_j \rho)=0$. In other words, the jump probability in the redefined Lindbladian equation vanishes if the effective Hamiltonian is Hermitian, indicating a unitary evolution. 
\section{GNS constrcution}\label{app:gnsconstruction}
The Gelfand–Naimark–Segal or GNS construction\, \citep{bratelli}, is the construction of a Hilbert space description of this system, which involves first constructing a $C^*$ algebra on which the state is faithful, i.e., has a trivial kernel. This algebra $\mathcal{U}_\o$ is constructed by first identifying the kernel of the state 
\begin{equation}
K_\o=\{X\in\mathcal{A}\,  \text{ s.t. }\, \o(X^*X)=0\},    
\end{equation}
and ``quotienting out" the kernel from the algebra: $\mathcal{U}_\o:=\mathcal{A}/K_\o$. In other words, the elements of $\mathcal{U}_\o$ consist of the equivalence classes 
\begin{equation}
\psi_A=\{A+X\,  \text{ s.t. }\, X\in K_\o\}    
\end{equation}
One can see that the functional $\o$ is faithful on $\mathcal{U}_\o$. Moreover,  $\mathcal{U}_\o$ is a complex vector space, a property one can check by using the fact that $AX\in K_\o\, \forall\, X\in K_\o$, i.e. $K_\o$ is a left ideal of $\mathcal{A}$. In the next step of the GNS construction, the state functional $\o$ is used to define an inner product on the vector space $\mathcal{U}_\o$ via $\miniprod{\psi_A}{\psi_B}=\o(A^*B)$. With this added structure, $\mathcal{U}_\o$ becomes a pre-Hilbert space which can be completed to form a Hilbert space $\mathcal{H}_\o=\overline{\mathcal{U}}_\o$. The bar over the algebra denotes the completion with respect to the norm induced by the inner product. Therefore, the set of vectors $\{\psi_A\}$ is dense in $\h_\o$. If one defines a representation given by the following action on the states:  $\pi_\o(A)\psi_B=\psi_{AB}$, one can then derive the relation
\begin{equation}
\o(A^*B)=\miniprod{\psi_\mathbb{1}}{\pi_\o(A^*B)\psi_\mathbb{1}}    
\end{equation}
where $\mathbb{1}$ is the identity element of the algebra $\mathcal{A}$. Therefore, $\psi_\mathbb{1}$ is a cyclic vector of the representation $(\h_\o,\pi_\o)$. This construction gives us a faithful cyclic representation of the algebra $\mathcal{U}_\o$, with the representation space $\h_\o$, the map $\pi_\o$, and the cyclic vector $\psi_\mathbb{1}$, such that $\o(A)=\miniprod{\psi_\mathbb{1}}{\pi(A)\psi_\mathbb{1}}_\mathcal{H_\o}$. 
\subsection{Combining the two representations}
Given two systems $\{\mathcal{A}_1,\o_1\}$ and $\{ \mathcal{A}_2,\o_2\}$ and their respective representations $\{\h_{\o_1},\pi^1_{\o_1}\}$ and $\{\h_{\o_2},\pi^2_{\o_2}\}$, we will construct a larger representations space. We define the Kernel space of these states as $K_{\o_1}=\{X\in \mathcal{A}_1\, \text{and}\, \o(X^*X)=0\}$ and similarly $K_{\o_2}$, and use them to construct the Hilbert spaces $\h_{\o_1}=\overline{ \mathcal{A}_1/K_{\o_1}}$ and $\h_{\o_2}=\overline{ \mathcal{A}_2/K_{\o_2}}$. Define the product of these spaces  $\h_{\o_1}\odot\h_{\o_2}$ as a linear combination of the elements of the form $\psi_{A_i}^1\psi_{B_j}^2$\, \citep{turumaru} where we have $A_i\in\mathcal{A}_1$ and $B_j\in \mathcal{A}_2$. This forms a linear vector space with an inner product defined through the state $\o(A_iB_j)=\o_1(A_i)\o_2(B_j)$. Note that this is not a faithful state over  $\h_{\o_1}\odot\h_{\o_2}$. This is because the kernel of $\o$ cannot be constructed solely from the kernels $K_{\o_1}$ and $K_{\o_2}$. One, therefore, defines a Hilbert space called the direct product Hilbert space by quotienting out this kernel:  $\h_{\o_1}\otimes\h_{\o_2}=(\overline{\h_{\o_1}\odot\h_{\o_2})/K_\o}$. Under this equivalence, the elements of the form $\psi_{A_i}^1\psi_{B_j}^2$ are mapped to the elements of the form $\psi_{A_i}^1\otimes\psi_{B_j}^2\in\h_{\o_1}\otimes\h_{\o_2}$.

\section{Spin tomography}\label{spintomo}
We will outline the method to tomographically construct a single particle spin 1/2 system. Given eigenvectors $\ket{m}$ of $\sigma_z$ with eigenvalues $m\in\{\dfrac{1}{2},-\dfrac{1}{2}\}$, the probability of obtaining the result $m$ after measuring the spin $\sigma_z$ is given by $p(m|\sigma_z)=\bra{m}\rho\ket{m}$. Consider a unitary evolution of this state under $U=(\mathbb{1}-i\sigma_Y)/\sqrt{2}$, the subsequent measurement of spin in the $\sigma_3$ direction, the probability of getting the result $m$, given by $p(m|\hat{n}_1)=\bra{m}U^\dagger\rho U\ket{m}$ is the probability of getting the result $m$ after a $\sigma_1$ measurement on $\rho$. This is because $U\ket{m}$ is the eigenvector of $\sigma_x$ with eigenvalue $m$. In general, if one wants to measure the observable $\vec{\sigma}.\vec{n}$ where $\vec{n}=(\sin\theta\cos\phi,\sin\theta\sin\phi,\cos\theta)$, it is equivalent to a $\sigma_z$ measurement of the state $U\rho U^\dagger$ where $U=e^{-i(\theta/2)\sigma.\hat{n}_\perp}$ is unitary with $\hat{n}_\perp=(-\sin\phi,\cos\phi,0)$. This is possible due to the following identity: $e^{-i(\theta/2)\sigma.\hat{n}_\perp}\sigma_z e^{i(\theta/2)\sigma.\hat{n}_\perp}=\vec{\sigma}.\vec{n}$\, . 
\begin{figure}
    \centering
    \includegraphics[width=0.8\linewidth,trim=4 4 4 4,clip]{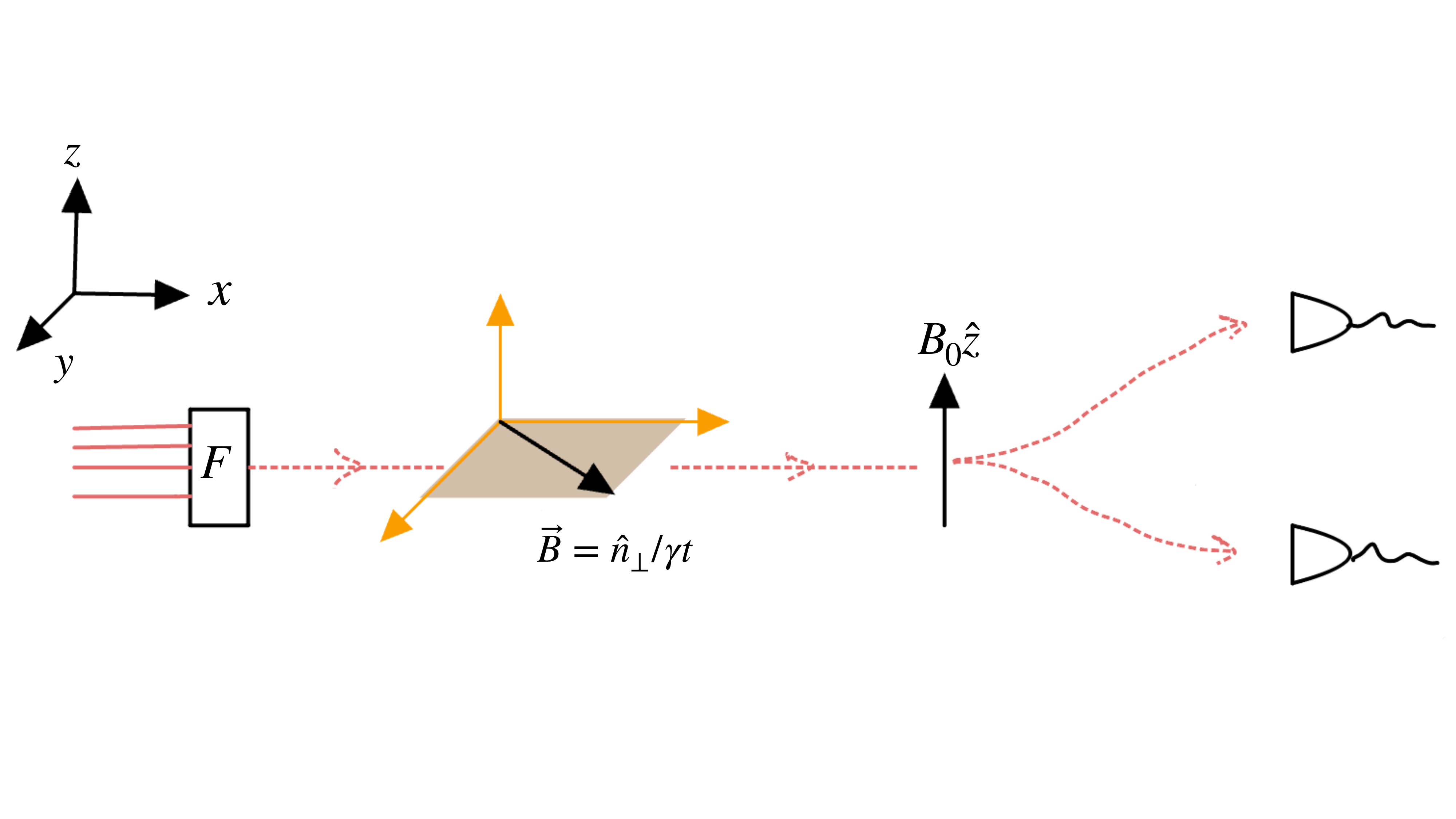}
    \caption{Spin tomography of a spin 1/2 particle. The filter $F$ lets particles of a certain velocity pass, which are then unitarily rotated by the magnetic field in the x-y plane. As they pass through the second magnetic field in z-direction, the two arms are detected, corresponding to $m=\pm1/2$ values.}
    \label{fig:spintomo}
\end{figure}
In the stern Gerlach setting, as shown in figure\ \ref{fig:spintomo}, a beam of electrons passes through a Fizeau filter, which allows particles with a specific velocity to pass through. This is required because the unitary rotation $U$ depends on the time for which the particle interacts with the magnetic field $\vec{B}$. The interaction Hamiltonian is given by $H=-\vec{\mu}.\vec{B}$, where $\vec{\mu}=\gamma\vec{\sigma}$ is the magnetic moment, $\gamma$ is the gyromagnetic factor and $\vec{B}$ is the magnetic field. The evolution $U$ is carried out by fixing this evolution is carried out by the magnetic field $\vec{B}=\theta\hat{n}_\perp/(\gamma t)$. After this, the particle passes through a magnetic field in $z$ direction which splits the beam into two, interpreted as $\pm1/2$ detection.
\\ \\
Given the data $\{m,p(m|\hat{n}_i)\}$, one can construct the density matrix using the Born rule. The probabilities are related to the state in the following way: $p(m|n_3)=\bra{m}\rho \ket{m}$, such that $\sum mp(m|n_3)=\dfrac{1}{2}\tr(\rho\sigma_3)$ is the expectation value of the $\sigma_3$ measurement. Fixing the magnetic field such that $\gamma Bt\vec{\sigma}.\hat{n}=\pi/4\sigma_1$\, and then passing the particle through a secondary magnetic field in $z-$direction, one measures $\sigma_2$. The probabilities are given by $p(m|\hat{n}_1)=\bra{m}U_1\rho U_1^\dagger \ket{m}$ where $U_1=e^{-i\pi/4\sigma_2}$. Similarly, for $U_2=e^{i\pi/4\sigma_1}$, we have $p(m|\hat{n}_2)=\bra{m}U_2\rho U_2^\dagger \ket{m}$. 
\end{document}